\documentclass[prl,twocolumn]{revtex4}
\usepackage{amsmath,amssymb,graphicx,psfrag,color}

\newcommand{\GUT}{{\rm GUT}}
\newcommand{\diag}{\text{diag}}

\begin{document}

\title{Contorted Flavors in Grand Unification and Proton Decay}
\author{Kang-Sin Choi}
\email{kschoi@th.physik.uni-bonn.de}
\affiliation{Physikalisches Institut, Universit\"at Bonn, Nussallee
12, D-53115 Bonn, Germany}

\begin{abstract}
Conventionally we collect electron and up/down-quarks
$(u,d)\oplus(\nu_e,e)$ to form unified generations,
and collect the left-handed and the right-handed fermions of the same
flavors $(\nu_e,e)\oplus \bar e$. We can alternatively have a
contorted multiplet, made by pairing different quarks and leptons, like
$(u,d)\oplus(\nu_\mu,\mu)$ or $(u,d)\oplus(\nu_\tau,\tau)$, and/or
between different left and right handed fermions to
$(\nu_e,e) \oplus \bar \mu$, etc. These can suppress
proton decay, due to its high flavor dependence, while
having the correct fermion masses. 
\end{abstract}

\maketitle

Unification of apparently different forces of Nature into a simpler
one has driven the history of physics. With the unification of weak
and the electromagnetic forces, the establishment of the Standard
Model of particle physics (SM) is based on the gauge theory of the strong
and the electroweak interactions described by the group $SU(3)_C
\times SU(2)_L \times U(1)_Y$. A further unification
was considered and resulted in Grand Unified Theory (GUT)
\cite{Georgi:1974sy,Pati:1973rp,so10}, possibly with the aid of the
supersymmetry (SUSY), which have been the best hint for the possible single
nature of strong and electroweak forces. Being gauge interaction,
extending the gauge groups into a (semi-)simple embracing group
naturally describes a unified force. For the size of unification
scale, the running of gauge coupling is the most important hint on
the gauge sector \cite{GQW}. Besides the
Grand Unification scheme at the field theory level, every known top-down
approach, for instance string theory, has the same structure of
a larger simple group and larger representations \cite{string}. Thus the
embeddability of quantum numbers are quite compelling.

Consequently the transforming matters are also unified to yield larger one(s).
Thus the extension fixes relative identities between the quark and the
lepton families. 
Upon forming unified multiplets, conventionally the following is assumed:
\begin{enumerate}
 \item Quarks (color triplets) and leptons (color singlets) form the
  same unified generations according to increasing order of masses.
 \item Left-handed (weak doublet) and right-handed (weak singlet)
  fermions of the same flavor form the same unified generations.
\end{enumerate}
Although conventional, there is {\em no {\it a
priori} reason to follow these assumptions.}
By relaxing the first assumption we can consider
\begin{equation} \label{twistgen}
 \begin{pmatrix} \nu_e \\ e  \end{pmatrix} \oplus
  \begin{pmatrix} u \\ d \end{pmatrix}
 \to \begin{pmatrix} \nu_\mu \\ \mu  \end{pmatrix} \oplus
  \begin{pmatrix} u \\ d \end{pmatrix}
 \text{ or } \begin{pmatrix} \nu_\tau \\ \tau  \end{pmatrix} \oplus
  \begin{pmatrix} u \\ d \end{pmatrix},
\end{equation}
and so on. For example, in one `contorted' version of 
$SU(5)$ model may contain $\overline{\bf 5}=(\bar
d_i, \nu_\tau, \tau)$\footnote{All the fermions here are
  left-handed Weyl spinors, with the charge conjugations for the
  barred letters. $i$ is color index.}.
By relaxing the second assumption we can have
\begin{equation} \label{twistlr}
 \begin{pmatrix} \nu_e \\ e \end{pmatrix} \oplus \bar e \to 
 \begin{pmatrix} \nu_e \\ e \end{pmatrix} \oplus \bar \mu,
\end{equation}
and so on.
In an extreme case, one unified generation can mix all the quarks and
leptons ${\bf \overline 5}=(\bar d_i, \nu_\mu, \tau), {\bf
  10}=(\begin{smallmatrix} c_i \\ s_i \end{smallmatrix}, \bar t_i, \bar e )$,
for instance.  
No physical principle restricted such pairing, such as anomaly
cancellation, showing each generation is completely closed even if we
consider different pairing. 
For a formation as in (\ref{twistgen}), we will see that the mass
hierarchy can be a guideline.

\section{Flavor structure}

After electroweak symmetry breaking, we diagonalize the flavor
eigenstates to the mass eigenstates
\begin{equation} \begin{split}
 \psi_F &\to U_F^\dagger \psi_F, \quad \bar \psi_F \to W_F \bar
 \psi_F, \\ 
 Y_F &\to Y_F^{\text{diag}}=U_F Y_F W_F^\dagger, \quad F=U,D,N,E,
\end{split} \end{equation}
with obvious notations \footnote{We will use upper case subscipts
  for matrices, and lower cases for their components. For
  example, $Y_e^\diag$ is the (11)-component of $Y_E^{\diag}$ and $V_{u\bar
  e}$ is (11)-component of the matrix $V_{U \bar E} \equiv 
  U_E^\dagger W_L$, etc.}.
In SM, we {\em define} the electron as the lightest charged lepton,
thus its flavor and mass eigenstates are identical. At the same time we 
independently define the up quark as the lightest quark,
because it is not connected to leptons by any interactions.
One may have equal right to define the muon as the lightest lepton,
in SM this is just renaming. However in GUT, the quarks
and leptons are related, for instance $e$ is related to
$u,\bar u$ in ${\bf 10}$ of $SU(5)$. Once we define the electron by fixing
$e$ and $\bar e$ in the entire multiplets $\bf 10, \overline 5$,
the up-type quarks follow from the relative mass
eigenstates $u',\bar u'$ by
\begin{equation} \label{eu}
 u = W_L U_U^\dagger u', \quad \bar u = U_L^\dagger W_U \bar u', \quad
 \text{with respect to } \bar e,
\end{equation}
to form the flavor multiplet. 
Therefore defining the muon as the lightest lepton modifies the
identity of the up quark as the lightest, since the new definition is now
distinguished. 
Thus in general, even the flavor eigenstates of up-type quarks are
not guaranteed to be mass eigenstates.
Also it means that we can distinguish contorted flavors, as in
(\ref{twistgen}) and (\ref{twistlr}). 
On the other hand, this information is not relevant in the low energy physics,
since there is no interaction connecting like (\ref{eu}) in SM. 
Even if a low level theory predicts the lighter mass for the muon than the
electron belonging to $(\begin{smallmatrix}u^i \\ d^i\end{smallmatrix}
  \bar u \ \bar e)$,
misidentification of the lightest as electron {\em does not} change SM.

After fixing the electron, the relative fermions in $SU(5)$ are
completely connected by the relative basis change matrices (RBCMs), among
them are CKM matrix $V_{UD} \equiv U_U^\dagger U_D$ \cite{CKM} and PMNS matrix
$V_{EN} \equiv U_E^\dagger U_N$ \cite{PMNS}, up to
renormalization effects. For example, the
emission of $Y^{1/3}$ boson from down quark to {\em 
conjugate}-up quark accompanies the RBCM $V_{d \bar u} \equiv (U_D^\dagger
W_U)_{d \bar u}$, which is to be distinguished from the CKM element $V_{d u}
\equiv (U_D^\dagger U_U)_{du}$. 

In the minimal $SU(5)$, we have two Yukawa couplings $Y_U = Y_U^T$ and
$Y_D = Y_E^T$  
diagonalized by four matrices $U_U, W_U, U_D = U_E, W_D = W_E$.
We can fix $U_U = V_{UD} =
V_{CKM}$ and $W_U = V_{CKM}^T P^T$, where $P$ is a diagonal $SU(3)$
matrix. Thus, besides known quark and lepton masses and CKM
matrix, we have two new phases \cite{JM}.
Since $SO(10)$ GUT completely relates all the fields in SM, we can in
principle completely observe the Yukawa coupling itself, from
interactions in GUT. This means, it is not possible to
arbitrarily align the flavor and the mass eigenstates.

\section{Contorted Flavors}

In the minimal $SU(5)$, the RBCM between down-type quarks and conjugate
leptons is $V_{D \bar E} = U_D^\dagger W_E = \bf 1$, without loss of
generality. It means that, for example by charged 
$X^{4/3}$ boson exchange, $d$ always transits to $\bar e$,
$V_{d \bar e} =1$, but there is no 
transition between $d$ and $\bar \mu$, $V_{d \bar \mu} =0$. 
It is because there is no difference in mass diagonalization between
down-type quarks and leptons $U_E = U_D, W_E = W_D$.

However the minimal models lead to bad mass relations $m_e / m_\mu = m_d/
m_s$, which is renormalization group invariant.  
One should introduce more Higgs fields and as many complex Yukawa
matrices. 
Usual fermion textures in non-minimal GUT imply similar RBCMs between
quarks and leptons, with slightly modified diagonality, but in general
there emerges many new phases.

To correct the relations, the main paradigm after Ref. \cite{GJ} has
been to introduce 45 and and higher dimensional Higgses. For
illustration, we adopt the most general renormalizable Yukawa coupling
including $\bf 45$:
$
 {\bf 10} \cdot {\bf 10} \cdot {\bf 5}_H + {\bf 10} \cdot {\bf
 \overline 5} \cdot {\bf
 \overline 5}_H +  {\bf 10} \cdot {\bf \overline 5} \cdot {\bf
 45}_H,
$
where the coefficients are in general complex matrices in the flavor
basis.
With some assumptions on the form of Yukawa couplings and VEV, we can
have the following form of mass matrices for isospin $-1/2$ fermions
\begin{equation}
  M_d = \begin{pmatrix} l-m & a & 0 \\ a & k+c & 0 \\ 0 & 0 & b
  \end{pmatrix}, \
  M_l = \begin{pmatrix} l+3m & a & 0 \\ a & k-3c & 0 \\ 0 & 0 & b
  \end{pmatrix}, \end{equation}
which is also a typical texture from `minimal renormalizable'
$SO(10)$, making use of $\bf 10$ and $\bf 126$ Higgses \cite{BM}.
The elements in the same letters come from the same VEVs of
$\bf \bar 5$ and $\bf 45$, with some Clebsch-Gordan coefficients.
If we take $l \sim m \sim k \sim 0$, in the parameter range $|a| \ll
|c| \ll |b|$, we have mass relations $m_d \simeq 3 m_e, m_s \simeq m_\mu/3, m_b
\simeq m_\tau$ after diagonalization \cite{GJ}. However in the
parameter range $|l - m| \ll 
|c+k|$ and $|k-3c| \ll |l+3m|$, with small $a$ and large $b$ as before, we have
{\em inverted lepton mass hierarchy} with respect to quarks. 
\begin{equation}
  m_d < m_s < m_b, \quad \text{`$m_\mu$'} < \text{`$m_e$'} < m_\tau.
\end{equation}
However we note that, although lepton mass hierarchy is inverted,
still {\em we identify the electron as the lightest} charged
lepton in SM. Therefore it will be more appropriate to collect the multiplet as
$(u,d)\oplus \mu$ and $(c,s)\oplus e$.
If the parameters are aligned to fit the observed fermion masses
$a^2/(l+3m) \simeq 0.5$ MeV$/c^2$, $l+3m \simeq 106$ MeV$/c^2$
(neglecting renormalization corrections), we cannot see the change in SM.

Of course, this has distinguished consequences in GUT, in particular
proton decay that we see shortly. 
Because we have inverted hierarchy of the charged leptons,
we have {\em mostly off-diagonal $U_E$ and $W_E$} with respect to the one
with normal hierarchy, 
\begin{equation} \begin{split}  
 U_E \to U_E T_{12},&\quad  W_E \to W_E T_{12}, \quad Y_E^\diag \to
 T_{12} Y_E^\diag  T_{12}^\dagger, \\
 &T_{12} = \begin{pmatrix}0&-1 & 0\\1&0&0 \\0&0&1
\end{pmatrix} \in S_3 \subset SO(3). 
\end{split} \label{rediag} \end{equation}
Now there is no transition between $d$ and $\bar e$, $V_{d \bar e}
\sim 0$ by $X^{4/3}$ boson exchange, but there is 
one between $d$ and $\bar \mu$, $V_{d \bar \mu} \sim 1$.
In more general contortions, $T$ can be any element of permutation
symmetry of order 3, $S_3$, which is a discrete subgroup of $SO(3)$.

In string compactifications, gauge bosons and matter fermions have
similar embedding as GUT \cite{Choi:2004vb,string}. However we do not
break them by Higgs 
mechanism, but by projection conditions associated with
symmetries of internal manifold, just leaving footprints of unified multiplets.
The fermion mass hierarchy is determined by geometry \cite{Choi:2007nb}, from
distributions of fields in the internal spaces. Thus we can have large
suppression for down-type quarks while having small suppression for 
leptons in the same multiplet and vice versa. Thus the contortion
naturally arises in string theory.

Because of symmetric Yukawa coupling, we cannot contort $u$ sector with
respect to $\bar u$ in $SU(5)$. Also $u$ and $d$ sectors are related
by CKM matrix, so we have a similar hierarchy between up and down type quarks.
We can further generalize the idea, as in (\ref{twistlr}),  by
exchanging {\em only the right-handed} fermions between different generations,
$ \begin{pmatrix} \nu_e \\ e  \end{pmatrix} \oplus
  \begin{pmatrix} u \\ d \end{pmatrix} \oplus \bar \mu
  \text{ and }
 \begin{pmatrix} \nu_\mu \\ \mu  \end{pmatrix} \oplus
  \begin{pmatrix} u \\ d \end{pmatrix} \oplus \bar e,
$
or
$ \begin{pmatrix} u \\ d \end{pmatrix} \oplus \bar t \oplus \bar d
  \text{ and }
  \begin{pmatrix} t \\ b \end{pmatrix} \oplus \bar u \oplus \bar b.
$
Then we have more freedom for quark and lepton masses. In
particular in $SU(5)$, highly asymmetric texture is possible,
thus the `minimal' $SU(5)$ can be different.
For example if we exchange $\bar u$ and $\bar c$ to form a contorted
family, the original symmetry constraint from ${\bf 10}_i \cdot {\bf
  10}_j$ on $Y_{u\bar c} = Y_{c\bar u}$ becomes
\begin{equation} 
 Y_{u \bar u} = Y_{c \bar c}.
\end{equation}

\section{Proton decay by dimension six operators}

For the association of quark and lepton
generations, the only direct experiment have been the proton decay
\cite{Kearns}.
First we consider the effect of dimension six operators.
The typical proton decay amplitude is $p \to e^+ \pi^0$
mediated by $X,Y$ bosons \cite{Hisano:2000dg}, 
\begin{equation} \label{pdecaywid} \begin{split}
 \Gamma=&\alpha^2 {m_p \over 64\pi f_\pi^2}(1+D+F)^2
 \left({g_5^2 A_R \over M_{\GUT}^2}\right)^2 \\
 &\times \Big(|V_{u\bar u}V_{e\bar d}^*|^2+
 |V_{u\bar u}V_{d \bar e}^*+V_{d\bar u}V_{u\bar e}^*|^2\Big).
\end{split} \end{equation}
Here $\alpha$ is the low energy contribution
from hadronic matrix element, can be calculated by lattice QCD
\cite{Aoki:2008ku}. $m_p,f_\pi,g_5,A_R$ are respectively the proton
mass, the pion decay constant, the unified gauge coupling, the
renormalization factor. $D 
\simeq 0.76$ and $F \simeq 0.48$ are interaction strength between
baryons and mesons obtained from chiral perturbation theory.
Neglecting the last factor, the partial lifetime for SUSY case is 
\begin{equation} \begin{split}
 \tau&/B = \Gamma^{-1} 
\simeq 8 \times 10^{34}\text{ years } \times \\
 &\left(0.015\text{GeV}^3 \over \alpha\right)^2\left({M_{\GUT} \over
 10^{16}\text{GeV}} \right)^4 \times
 \text{(flavor factor)}.
\end{split} 
\end{equation}
The gauge boson mass is guided by running gauge coupling:
$M_{\GUT} \sim 10^{15}$ GeV for non-SUSY and $M_{\GUT} \simeq
3\times 10^{16}$ GeV for SUSY cases \cite{gutsrpp}.
Considering a typical decay channel like $p \to e^+ \pi^0$,
we have the partial mean lifetime, in non-SUSY case
$\tau/B \sim 10^{31\text{-}35}$ years, depending on extensions.
The observed bound for the proton lifetime is $\tau/B > 5 \times
10^{33}$ years \cite{gutsrpp}.

The last factor contains RBCMs,
which are respectively the components of $U_U^\dagger W_U, (U_E^\dagger
W_D)^\dagger, U_U^\dagger W_U, (U_D^\dagger W_E)^\dagger, U_D^\dagger
W_U, (U_U^\dagger W_E)^\dagger$, all of which lie outside CKM and PMNS.
The amplitude is highly flavor dependent \cite{FileviezPerez:2004hn,JM}.
In the minimal $SU(5)$, all the elements are set to identity as above, except
$V_{U \bar U} = P_{CKM} P, V_{D \bar U} = V_{CKM}^T P^T, V_{U \bar E}
= V_{CKM}^T P^T V_{CKM},$ so the flavor factor is nearly 5. 
The flipped $SU(5)$ \cite{flippedsu5} contains only the first term in the
last factor \cite{Dorsner:2006ye}.

If we contort the first two generations as in (\ref{rediag}),
then we have $V_{Q\bar L} \simeq T$ with $V_{u,d\bar e} \sim 0, V_{u,d
  \bar \mu} \sim 1$. Thus the transition from quark to lepton is
  dominated by off-diagonal elements. Eventually the proton decay 
 is dominated by muon. 
Consequently from (\ref{pdecaywid}) there can be no $p \to e^+ \pi^0$,
but {\em opens proton decay channels to other leptons,} like
\begin{equation}
 p \to \mu^+ \pi^0,
\end{equation}
with the same lifetime as the original case.
The bound for this process is $\tau/B > 3.7 \times 10^{33}$ years, is
almost same order of magnitude to proton decay into positron. Thus
large off-diagonal RBCM element does not help much.

There have been only water (e.g. Super-Kamiokande) and earthly metal
(e.g. iron in Soudan II) based experiments, thus are able to
identify decay from protons and neutrons, consisting of up and down quarks
as initial states. So if the electron is connected with other quarks
than $u,d$, we cannot observe the decay from other sources.

The pairing $(u,d)\oplus\tau$ is achieved if we invert
the first and the third generation leptons.
The outgoing state would be $u d \to \tau^+ \bar u$, instead of $u
d \to e^+ \bar u$. In this case
the proton cannot decay into tauon, since it is heavier ($m_\tau=1777$
MeV$/c^2$) than proton ($m_p=938$ MeV$/c^2$). Thus if the tauon
component of the mixing is high, we will lose decay information for $p
\to e^+ \pi^0$. However still there are decay channels
\begin{align}
 p & \to \bar \nu \pi^+ & \tau/B > 2.5 \times 10^{31} \text{
   years}, \label{ptonupi} \\
 n & \to \bar \nu \pi^0 & \tau/B > 1.1 \times 10^{32} \text{
   years}, \label{ntonupi}
\end{align}
where we cannot distinguish the flavor of neutrino, since we only seek
missing energy. Notably this decay does not exist in the flipped
$SU(5)$. 

In some model, with contortion like (\ref{twistlr}), we can suppress
some nucleon decay by dimension six operators, however it is very hard
to overcome the constraint from (\ref{ptonupi}) and (\ref{ntonupi}).
It is because, exchanging right-handed quarks and leptons, there is
always a pair $\overline u \oplus e$ in almost all known unification
models. Thus a nucleon decay into
neutrino is very important verification of all kind of GUT
interactions \cite{Dorsner:2004xa}.

We may make theory with just simple exchange $e \leftrightarrow
\mu$ to have a generation $(u,d) \oplus \mu$, without mass
inversion. Then we can easily see that, since the RBCM does not change
$V_E = W_E = {\bf 1}$, so that proton always decays into the lightest lepton,
whatever name it has. Thus we have no change.

\section{Proton decay by dimension five operators}

\begin{figure}
\psfrag{f1}{$\bar u$}
\psfrag{f2}{$\bar d$}
\psfrag{v1}{$ Y^\diag_d \color{red} V_{\bar d \bar u}$}
\psfrag{v2}{$T$}
\psfrag{v3}{$Y^\diag_\tau V_{\nu_\tau \tau}$}
\psfrag{v4}{$\color{blue} Y_{t,c}^\diag V_{t,c s}$}
\psfrag{v5}{$M_T Y^\diag_t \color{red} V_{\bar t,\bar c \bar \tau}^*$}
\psfrag{v6}{$\mu$}
\psfrag{f3}{$\nu_{\tau}$}
\psfrag{f4}{$s$}
\psfrag{f5}{$\tilde{\bar \tau}$}
\psfrag{f6}{$\color{blue} \tilde{\bar t}, \tilde{\bar c}$}
\begin{center}
\includegraphics[height=3.5cm]{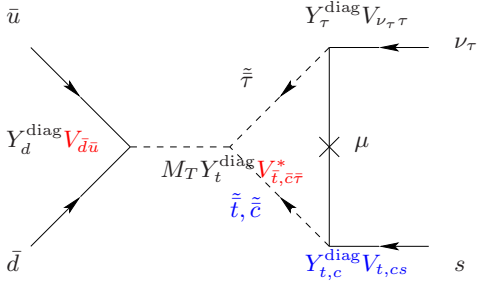}
\end{center}
\caption{\label{fig:r4} The most dominant dimension five proton decay
  operators of $RRRR$ type in SUSY $SU(5)$ GUT.}
\end{figure}

Although the dimension six operators are not so harmful in SUSY GUT, there
are relevant dimension five operators from integrating out the triplet
Higgs pair, known as $L^4$ and $R^4$:
$c_{5L}^{ijkl}Q_i Q_j Q_k L_l /(2M_T)+
c_{5R}^{ijkl} \bar u_i \bar d_j \bar u_k \bar e_l /M_T,$ where
\begin{align}
   c_{5L}^{ijkl} = (Y_D^\diag)^{km} (V_{Q L})^l_m
   (Y_U^\diag)^{in} (V^\dagger_{Q Q})^j_n, \\ 
   c_{5R}^{ijkl} = (Y_D^\diag)^{mj} (V_{\bar D \bar U})^i_m
   (Y_U^\diag)^{nl} (V^\dagger_{\bar U \bar E})^k_n,
\end{align}
at the unification scale.
Here $M_T$ is the triplet Higgs mass, obtained by unification condition.
Because of $SU(3)_C\times SU(2)_R$ contractions, $c_{5L}$ vanishes if
the family indices satisfy $i=j=k$ and $c_{5R}$ vanishes if $i=k$.
Considering all the contribution, the
dominant decay channels are $p \to \bar \nu_\mu K^+$ for $L^4$, a
`box' diagram dressed with wino loop, and $p \to \bar \nu_\tau K^+$ for $R^4$ 
with Higgsino loop, depicted in Fig. \ref{fig:r4}. The latter has amplitude  
\begin{equation}
 A = [A_\tau(\tilde t_L) + A_\tau(\tilde c_L)]_{L^4} + [A_\tau(\tilde t_R)
   + A_\tau(\tilde c_R)]_{R^4},
\end{equation}
where
\begin{align}
 A_\tau(\tilde c_L) & \simeq c_{5L}^{1123} g^2_2 M_2/(M_T m_{\tilde
   f}^2), \\
 A_\tau(\tilde t_L) & \simeq c_{5L}^{1133} g^2_2 M_2/(M_T m_{\tilde
   f}^2),\\
 A_\tau(\tilde c_R) & \simeq c_{5R}^{1123} Y_\tau V_{\nu \tau}^* Y_c
 V_{cs} \mu / (M_T  m_{\tilde f}^2), \label{5rscharm}\\
 A_\tau(\tilde t_R) & \simeq c_{5R}^{1133} Y_\tau V_{\nu \tau}^* Y_t
 V_{ts} \mu / (M_T  m_{\tilde f}^2), \label{5rstop}
\end{align}
where $g_2, M_2$ and $\mu$ are weak coupling, wino mass and mu
parameter, respectively.
The experimental bound is $\tau/B > 6.7 \times 10^{32}$ years, which is
blind to neutrino species, so it includes all the species.

In the minimal $SU(5)$, $V_{ql}$ and $V_{\bar u \bar e}$
contain the same phase matrix $P$ which belongs to a diagonal $SU(3)$.
It is well known that the $L^4$ operators can be suppressed by
destructive interference between two amplitudes mediated by scharm
$\tilde c_L$ and stop $\tilde t_L$, adjusting phases
$\phi_{23} \equiv -i\log (P_{22}/P_{33}) \simeq 160^\circ$ \cite{Nath:1985ub}. 
Also it is calculated that the $R^4$ operators for the tau  neutrino
decay contributes much larger, enhanced by tau and top Yukawa
couplings and the mu parameter \cite{GN}. There the term (\ref{5rscharm}) with
right-scharm loop was neglected, due to Yukawa suppression by
$(Y_c/Y_t)^2$ in the amplitude. In fact the suppression power is closer
to 1, since we have additional CKM factors to have $Y_t V_{ts}
\sim Y_c V_{cs}$.  In the minimal $SU(5)$, both $c_R^{1123}$ and
$c_R^{1133}$ have the same phase $P_{11}$.

We note that in the minimal $SU(5)$ it is implicitly used that
$V_{\bar t \bar \tau} \simeq 1$. In a
contorted model, if it became small, the stop exchange
amplitude could be comparable to that with scharm loop.
This can be done by exchanging tauon with muon or electron, as in
(\ref{twistgen}),  because we then have $W_l \simeq T$ where $T$ is
now mostly off-diagonal in (23), or (13) components, as in
(\ref{rediag}). 
Considering the former case $T=T_{23}$, from  $V_{\bar u \bar l} =
V_{CKM}^T T_{23} P$ we have $V_{\bar t \bar \tau} \simeq V_{cb} =
0.04, V_{\bar c \bar \tau} \simeq V_{tb} \simeq 1$ so that 
\begin{equation}
  Y_t^2 V_{\bar t \bar \tau}^* V_{ts} \sim Y_c^2 V_{\bar c \bar \tau}^* V_{cs}.
\end{equation}
Then an additional phase in the RBCM $V_{\bar u \bar l}$ would help to cancel
$R^4$ part. Besides this effect, a contortion on the lepton sector
does not affect other vertices of the proton decay operator, except
the common part $V_{\nu_\tau \tau}$. In fact this part quite insensitive to
our contortion because of large mixing of PMNS matrix.
 In the latter case $T=T_{13}$, we have $V_{\bar t
  \bar \tau} \simeq V_{ub} = 0.004$ and $V_{\bar c \bar \tau} \simeq
V_{cb} = 0.04$ thus the stop and scharm loop contributions are smaller
and comparable. 
In any case there must be enhancement in operators including the
electron or the muon, since now either $V_{\bar t \bar e} \sim 1$
or $V_{\bar t \bar \mu} \sim 1$. Thus we cannot suppress all the 
$R^4$ operator contribution simultaneously.
In the case of twisting between electron and tauon, with
the dominant channel $p \to \bar \nu_e K^+$, we can
enhance the proton lifetime by the factor $(Y_\tau m_{\tilde e}/Y_e
m_{\tilde \tau})^2$
with respect to (\ref{5rstop}). This restores minimum peak of the $p \to
\bar \nu_\tau K^+$ back to $\phi_{23} \simeq 160^\circ$, while pushing $p
\to \bar \nu_e K^+$ to higher $\phi_{23}$. There are no sizable
enhancements in the other box diagrams due to small Yukawa couplings
$Y_e^{\diag}$ or $Y_u^{\diag}$. 

\acknowledgments
The author is grateful to Mitsuru Kakizaki, Bumseok
Kyae for discussions.
This work is supported by the DFG cluster of excellence Origin and
Structure of the Universe, the European Union 6th framework program
MRTN-CT-2004-503069 ``Quest for unification", MRTN-CT-2004-005104
``ForcesUniverse", MRTN-CT-2006-035863 ``UniverseNet" and
SFB-Transregios 27 ``Neutrinos and Beyond" and 33 ``The Dark 
Universe" by Deutsche Forschungsgemeinschaft (DFG).

\end{document}